\documentstyle[12pt]{article}

\title{ Multicomponent nonisothermal nucleation. 2. Chapman-Enskog procedure}

\author{V.B.Kurasov}

\date{Victor.Kurasov@pobox.spbu.ru}

\begin{document}

\maketitle

This part of the theory directly continues the derivation of the kinetic
equation presented in the preprint 990958 of this archive. All definitions
are the same and no special remarks are given.

\section{Chapman-Enskog procedure}

At first we recall the standard version of the Chapman-Enskog procedure
and then we shall present the generalization to our  situation  and
fulfill
calculations.

Consider equation
\begin{equation}\label{1}
\frac{\partial P}{\partial t} =
- (A+B) P
\end{equation}
where operators $A$ and $B$ have the following properties
\begin{itemize}
\item
Operator $A$ depends on variable $x$, operator $B$ depends on $x$ and
$y$. Both operators don't depend on time $t$.
\item
Operator $B$ is supposed to be small in comparison with $A$. Later we
shall rescale time as to have $|| A || \sim 1$, $|| B || \ll 1$. The value
$|| B || $ will be the small parameter of the theory.
\item
The eigenvalues and eigenfunctions of $B$ are unknown. The eigenfunctions
$A_i $ and eigenvalues $a_i$ of operator $A$
are supposed to be known. They have the following
properties
$$
a_0 = 0
$$
$$
a_i > \delta  > 0 \ \ \ \ i \ne 0
$$
with some positive small $delta$.
The eigenfunctions are supposed to be normalized.
\end{itemize}

One can introduce symbol $O(||B||)$ which means that the given expression
has the order of $||B||$ or less.

Initial conditions for     (\ref{1}) are supposed to be known.
Then one can directly describe the evolution during the first periods
of time. Here one can neglect $B$. One has only to solve equation
\begin{equation}
\frac{\partial P}{\partial t} =
- A P
\end{equation}
Due to the known eigenfunctions of $A$ this solution is quite obvious
\begin{equation}                     \label{2}
P = \sum_{m=0}^{\infty} \exp(-a_m t) p_m A_m
\end{equation}
where
\begin{equation}
p_m = (P|_{t=0} , A_m)
\end{equation}
will be called the mode amplitudes ($A_m$ will be modes) and the symbol
$( \  , \  )$ denotes the scalar product.

The final of relaxation according to          (\ref{2}) is obvious
\begin{equation}
P \rightarrow  p_0 A_0
\end{equation}
The characteristic time of relaxation is
\begin{equation}
t_{rel} = 1/a_1
\end{equation}
where $a_1$ is the  smallest eigenvalue of $A$ (except the zero value).

Now $A(p_0 A_0) = 0$ and operator $A$  exhausts its power. One has to
consider the action of $B$\footnote{
The physical reason to consider the further stages of evolution is that
in the nucleation problem $A_0 = 1$ and the final state of relaxation
doesn't provide ant benefit into the value of the nucleation rate. It
lies in the evident correspondence with the formula $J \sim \exp(F_c) /
\Delta \nu$ of the one-dimensional theory
where stands the big parameter $\Delta \nu$. }.

The deviation of $P$ from $p_0A_0$ can be caused only by the action of
operator $B$. So, in some sense it is small. Then we shall seek  the
solution in the following form
\begin{equation}
P = p_0 A_0 + \Delta
\end{equation}
For the correction term $\Delta$ in accordance with the perturbation theory
one can consider the representation
\begin{equation}
\Delta = \sum_i^{\infty} \delta^{(i)}
\end{equation}
where every term $\delta^{(i)}$ has the order $||B||^i$:
$$
\delta^{(i)}  \sim O(|| B ||^s)
$$

Now we are going to consider the time dependence of the amplitudes $p_m$.
One can note that the action of $A$ leads to the vanishing of $p_m$ for
$ m \ne
0$. The unique source to fill the given mode is to take into account the
action of operator $B$ which is small but still can provide the transition
from one mode to another. But $B$ doesn't depend on time. So, the intensities
of transitions between modes don't depend on time also. So, the unique
source of time dependence is the time dependence of the zero mode $p_0$.
Later this dependence will be spread by $B$ to all other modes.

As the result we state that the time dependence of an arbitrary mode amplitude
is going through the time dependence of $p_0$:
\begin{equation}
p_m(t) = p_m(p_0(t))
\end{equation}
\begin{equation}
\frac{\partial p_m}{\partial t} =
\frac{\partial p_m}{\partial p_0}
\frac{\partial p_0}{\partial t}
\end{equation}

Certainly the time dependence of $p_0$ isn't fixed and can not be obtained
immediately. One has to recalculate it at the every step of
approximation\footnote{It
is also possible to leave this dependence as some formal parameter and
to calculate it when the "final" approximation (i.e. the approximation
with the necessary precision) is obtained.}.
The value $\delta^{(i)}$ depends on time $t$ through $p_0(t)$ in the $i$-th
approximation.    At every new\footnote{The lower index denotes modes.}
 $\delta^{(i)}$ one has  new $p_m^{(i)}$
and new $p_0^{(i)}$. So, one has a new time dependence.

The substitution of expansion
\begin{equation}
P = p_0 A_0 + \sum_{i=1}^{\infty} \delta^{(i)}
\end{equation}
into (\ref{1}) gives
\begin{equation}
\frac{\partial p_0}{\partial t} A_0 +
\sum_{i=1}^{\infty} \frac{\partial \delta^{(i)} }{\partial t} =
- (A+B) (p_0 A_0 + \sum_{i=1}^{\infty} \delta^{(i)})
\end{equation}
Note that the $i$-th term has order $i$, but the value with the order $i$
will have benefits from all terms  with $j \geq i$.

The r.h.s. and l.h.s. of the last equation are some functions and we suppose
that they can be decomposed into series over the small parameter $|| B||$.
We shall denote the $i$-th term
which has the order $||B||^i$ by the superscript $<i>$. Then one can
write for the r.h.s.
\begin{equation}
 ( (A+B) (p_0 A_0 + \sum_{i=1}^{\infty} \delta^{(i)}))^{<i>} =
A \delta^{(i)} + B \delta^{(i-1)}
\end{equation}

The consideration of the l.h.s. is more complicated. At first we have
to see the expression for $\partial p_0 / \partial t$.  Having projected
equation (\ref{1}) on $A_0$ one can see that
\begin{equation}
\frac{\partial p_0}{\partial t} =
-(A_0 , B P)
\end{equation}
Hence, one can come to the estimate
$$
\frac{\partial p_0}{\partial t} = O(||B||)
$$
One can rewrite the last relation using the decomposition (\ref{2})
\begin{equation}  \label{kkk}
\frac{\partial p_0}{\partial t} =
-(A_0 , B( p_0 A_0 + \sum_{i=1}^{\infty} \delta^{(i)}) )
\end{equation}
Now it is evident that          $ {\partial p_0} / {\partial t}$
contains all powers of the small parameter and it has the order $O(||B||)$.
Now it is worth to determine the $j$-th term in  decomposition over parameter
$||B||$.  It quite easy to do for  ${\partial p_0} / {\partial t}$.
Namely,
\begin{equation}                  (
 \frac{\partial p_0}{\partial t} )^{<i>} =
- (A_0 , B \delta^{(i-1)} )
\end{equation}

Then one can fulfill the same transformations for $\partial \delta^{(i)}
/ \partial t$.  One present the following estimate
\begin{equation}
\frac{\partial \delta^{(i)} } { \partial t } =
\frac{\partial \delta^{(i)} } { \partial p_0 }
\frac{\partial p_0 } { \partial t } \sim
O(||B||^i) O(||B||) \sim O(||B||^{i+1})
\end{equation}
So, one can reduce the infinite sum to the finite sum
\begin{equation}
(\sum_{j=1}^{\infty} \frac{\partial \delta^{(j)} }{\partial t})^{<i>}
\rightarrow
(\sum_{j=1}^{i-1} \frac{\partial \delta^{(j)} }{\partial t})^{<i>}
\end{equation}
The next transformations are evident
\begin{eqnarray}
(\sum_{j=1}^{i-1} \frac{\partial \delta^{(j)} }{\partial t})^{<i>} =
(\sum_{j=1}^{i-1} \frac{\partial \delta^{(j)} }{\partial p_0}
\frac{\partial p_0 }{\partial t}
)^{<i>}
=
\\ \nonumber
\sum_{j=1}^{i-1}  \frac{\partial \delta^{(j)} }{\partial p_0}
(\frac{\partial p_0 }{\partial t})^{<i-j>}
=
\\ \nonumber
 - \sum_{j=1}^{i-1}  \frac{\partial \delta^{(j)} }{\partial p_0}
  (A_0 , B \delta^{(i-j-1)} )
\end{eqnarray}
$$
\delta^{(0)} = p_0 A_0
$$

Now we can substitute all results into the initial equation and get
\begin{eqnarray}
- (A_0 , B \delta^{(i-1)} ) A_0
- \sum_{j=1}^{i-1}  \frac{\partial \delta^{(j)} }{\partial p_0}
  (A_0 , B \delta^{(i-j-1)} )
=
\\ \nonumber
- A \delta^{(i)} - B \delta^{(i-1)}
\end{eqnarray}

As far as the eigenfunctions of $A$ are known  one can use the last relation
to get $\delta^{(i)}$.

\section{Relaxation}

Consider now our situation. Certainly, one can choose  operator $D_1$
as the main one.

An operator
$$
D_0 =
(\frac{\partial}{\partial \mu} - 2 \mu)
\frac{\partial}{\partial \mu}
$$
has the known set of eigenfunctions which are the Hermite polynomials
$H_j$ ($j=0,1,2,3,4,.....$). The scalar product is defined as
$$
(\Psi, \Phi) =
\frac{1}{\sqrt{\pi}} \int_{-\infty}^{\infty} d x  \exp(-x^2)
\Phi (x) \Psi (x)
$$
which leads to
$$
(H_j , H_i ) = \delta_{ij} 2^i i!
$$
The eigenvalues of $D_0$ are
$$
\lambda_i = -2i
\ \ i=0,1,2,...
$$
One has to mention two important relations
$$
(\frac{\partial}{\partial \mu} - 2 \mu) H_i =
- H_{i+1}
$$
$$
\frac{\partial}{\partial \mu} H_i = 2 j H_{j-1}
$$
The first one gives the way to construct the set of Hermite polynomials.
Both these equations allow to introduce in \cite{book} the representation
of modes.
We shall call the operators in the r.h.s. of two previous equations as
the transition operators. The first one is the mode increase operator.

The second one is the mode decrease operator. Then one  can
consider the operators of multiplication on $\mu$ and
the differentiation over $\mu$ as the superposition of transition operators.
Any additional operator appears from the non-Fokker-Planck behavior and
from higher terms in the free energy. Non-Fokker-Planck terms give high
derivatives. Free energy can be well approximated in the nearcritical
region by polynomial and leads to additional multiplications on $\mu$. So,
the additional term can be regarded as the superposition of the mode
increase operators and mode decrease operators.  This procedure resembles
the formalism of the secondary quantization. Below we shall use another
approach.

It is clear that $D_1$ can be presented as
$$
D_1 = \sum_{l=1}^{\infty}
d^{(1)}_l
(\frac{\partial}{\partial \mu} - 2 \mu)^l
\frac{\partial^l}{\partial \mu^l}
$$
with coefficients
$$
d^{(1)}_l =
-
\sum_j  \frac{(-\tau_j)^l}{l!}    S_{2j}
 W^+_j   \frac{\tau_j^l}{l!}
\ \ \ \  m=2,3,4,.....
$$
\begin{eqnarray}
d^{(1)}_1 =
\sum_{j'}
 W^+_{j'} \alpha_{acc\ j'} \frac{c_{g\ j'}} {2 \sum_j c_j \nu_j}
+
\nonumber
\\
\sum_{i'}
 W^+_{i'} (1-\alpha_{c\ i'}) \alpha_{acc\ i'} \frac{c_{i'}} {2 \sum_j c_j
\nu_j}
+
\nonumber
\\
\nonumber
\sum_{i'}
 W^+_{i'} \alpha_{c\ i'} \frac{c_{i'}} {2 \sum_j c_j \nu_j}
-
\sum_j  (-\tau_j^2)
 W^+_j S_{2j}
\end{eqnarray}
Then the eigenfunctions of $D_1$ will be the Hermite polynomials\footnote{In
the main order $S_{2j}$ goes to $1$.} and the
eigenvalues are given by
$$
\lambda_j =     \sum_{l=1}^j d^{(1)}_l (-2)^l \frac{j!}{(j-l+1)!}
$$

As the result it can be seen that one can take $D_1$ as the main operator
in the Chapman-Enskog procedure.
The approximate form of kinetic equation will be the following one
$$
\frac{\partial P}{\partial t}
 =
D_1 P
$$
But already in the investigation of the
stationary distribution establishing one will see that the formal priority
of $D_1$  in  comparison with $D_2$ isn't sufficient to ensure
the relaxation
to the stationary state.

Operator $D_2$ will be small in comparison with $D_1$ in terms of some
small parameter (not in formal sense described above) only when $\tau_j$
far all $j$ are small parameters. Then the main terms in operator $D_2$
will be the terms with the smallest sum of indexes $l+m$.  Then the
required
condition of the smallness of $D_2$ formulated in terms of the relaxation
times will be
$$
p_{rel} \equiv
\frac{
|
 - \sum_j \tau_j^3
 W^+_j
|
}{|\lambda_1|} \ll 1
$$
This parameter has been required\footnote{with account of the mentioned
error.} to be small in \cite{Djik}. Then the
solution at the relaxation stage can be written as the seria of the relaxation
modes
$$
P = \sum_{l=0}^{\infty} p_l^{(0)} \exp( - \lambda_l t)  H_l
$$
where
$$
p_l^{(0)} = (2^l l!)^{-1} (H_l , P(t=0) )
$$
are initial amplitudes.
This seria ensures the relaxation to the stationary state $p_0^{(0)} H_0$:
$$
P \rightarrow  p_0^{(0)} H_0 \equiv P_{rel}
$$
The time of relaxation $t_{rel}$
is given by
$$
t_{rel\ D_1} = \lambda_1
$$

The mentioned restriction $p_{rel} \le 1$ (actually according to \cite{Djik}
one has to require
$p_{rel} \ll 1$ )
practically excludes the nucleation under the strong thermal effects.
So, the situation considered in \cite{Djik} is rather poor.

One can note that the final state
$P_{rel}$ corresponds to the equilibrium distribution over $\mu$. So, it
is the eigenfunction of the operator $D_1 + D_2$ with the zero eigenvalue
$$
(D_1+D_2) P_{rel} = 0
$$
This can be easily proven when we return to the form of the finite differences
and reconstruct $D_1$ and $D_2$.

Certainly it is difficult to determine all eigenfunctions and eigenvalues
of $D_1 + D_2$.
We needn't all relaxation modes but only the final one. Also we have to
estimate the time of relaxation.
One can  prove that the relaxation time $t_{rel\ D_1+ D_2}$ for equation
$$
\frac{\partial P}{\partial t}
 =
(D_1+D_2) P
$$
is less or equal to the previous relaxation time $t_{rel\ D_1}$
$$
t_{rel\ D_1+D_2} \leq t_{rel \ D_1}
$$
Operator $D_2$ "helps" the relaxation to $P_{rel}$.
The way to prove the last estimate is to consider the blocks along $\mu$
axis and estimate  the action of $D_2$ between blocks.

As far as $D_3$,
$D_4$ have small parameter in comparison with $D_1$
then no special condition
is required. Now we can consider $D_1+D_2$ as the main operator
at the relaxation stage. The
smallness of $p_{rel}$ isn't now required.

\section{Correction terms}

When $P$ is close to $P_{rel}$ which is the eigenfunction with the zero
eigenvalue the operators $D_1$ or $D_1+D_2$ can not be considered as the
main ones. One has to use the Chapman-Enskog asymptotic decomposition.
But to fulfill this decomposition one has to know all eigenfunctions of
the main operator. So we can not use $D_1 + D_2$ as the main operator
now.

We shall redefine the main operator after the end of the relaxation
stage. Now the main operator will be $D_1$. One can see that
"correction operator"
$D_2+D_3+D_4$ has a complicate structure. Operators $D_3$ and $D_4$
have small parameter. Operator $D_2$ is small only in the formal sense.
So, the natural modification of the Chapman-Enskog procedure is to consider
two Chapman-Enskog procedures. One can include
the whole
internal  Chapman-Enskog procedure into every step of the external
Chapman-Enskog procedure. The initial procedure has the aim to "invert"
the action of operator $D_2$ and the external procedure will "invert" the
action of $D_3 + D_4$.

Every new approximation will contain the small parameter $1/\Delta \kappa
\sim \kappa^{2/3}$  ( or at least $\Delta \kappa / \kappa \sim \kappa^{-1/3}$)
in the order of the number of external approximation. This order decrease
the order of flow calculated in the initial (zero) approximation. But
one has to stop the external procedure until the resulting order is greater
than $\kappa^0 = 1$ because the continuous
description along $\nu_i$ will be violated in this order. One has to note
that the evolution along $\nu_i$ is principally different in comparison
with evolution along $\mu$. The domain along $\nu_i$ is concentrated
only at the integer values. The value of $\mu$ is the real value. This
difference leads to the specific lattice corrections described in \cite{book}.
These corrections have very complicated structure, an account can not
be fulfilled explicitly in the analytic way.  As the result we have to
restrict ourselves by the first correction and adopt that the account
of the first  correction in the external procedure is already sufficient.
But in the internal procedure we have to calculate the great number of
approximations.

Also one has to note here the requirement to have no operators $S_{1j}$,
$S_{@j}$, $S_{3j}$ in the final results. Really, the presence of this
operators corresponds to the elementary shift taken into account. This
lies in contradiction with the lattice domain along $\nu_i$.

The matter under discussion is the rate of nucleation and to determine
this rate and need only the embryos flows averaged  over $\mu$. This
operation corresponds to the projection on $H_0 \sim const$.

Now we can turn ourselves  to establish equations for $\partial p_0 / \partial
t$.
The last dependence is extremely important because in the Chapman-Enskog
procedure all dependence of approximations on time is going through
dependence
of $p_0$ on time. To get  $\partial p_0 / \partial t$ one can project
kinetic equation
$$
\frac{\partial P}{\partial t} = [ D_1 + D_2 + D_3 + D_4 ] P
$$
on $H_0$ i.e. fulfill the integration $\int_{-\infty}^{\infty} \exp(-\mu^2)
..... d\mu$. The l.h.s. gives $\partial p_0 / \partial t$. Then
$$
\frac{\partial p_0}{\partial t} =
(H_0, [D_1+D_2+D_3+D_4]P)
$$
The r.h.s.
will lead to more complicate expression. Consider
$$
(H_0, D_1 P) = (H_0, D_1 \sum_i p_i H_i) = (H_0, \sum_i \lambda_i p_i H_i)=
\lambda_0 p_0 = 0
$$
as far as $\lambda_0 = 0$.
Consider
$$
(H_0, D_2 P) =
(H_0, D_2 \sum_{i=0}^{\infty} p_i H_i )
$$
>From explicit form of $D_2$ one can see that the last elementary operator
in action of $D_2$ will be the mode increase operator
$(\frac{\partial}{\partial \mu } - 2 \mu )$. So, if we present
$
 D_2 \sum_{i=0}^{\infty} p_i H_i
$
as
$$
 D_2 \sum_{i=0}^{\infty} p_i H_i
=
\sum_{i=0}^{\infty} c_i H_i
$$
then
$c_0 = 0$. There is no projection on the zero mode. Then
$$
(H_0, \sum_{i=1}^{\infty} c_i H_i) = 0
$$
for arbitrary $c_i$. Then
$$
(H_0, D_2 P) = 0
$$

Consider $(H_0, D_3 P)$. Operator $D_3$ can be split into two parts
$D_{3a}$ and $D_{3b}$ defined as
$$
D_{3a} =
\sum_j \sum_{l=1}^{\infty} \frac{(-\tau_j)^l}{l!}
(\frac{\partial}{\partial \mu} - 2\mu)^l
L_j  S_{1j} P(\{\nu_i \},  \mu)
$$
$$
D_{3b} =
-
\sum_j
\frac{\partial}{\partial \nu_j }
 W^+_j S_{3j}  \sum_{m=1}^{\infty} \frac{\tau_j^m}{m!}
\frac{\partial^m}{\partial \mu^m} P(\{ \nu_i \}, \mu)
$$
$$
D_3 = D_{3a} + D_{3b}
$$
Then
$$
(H_0, D_{3a} P) = 0
$$
and the reasons are the same as for $D_2$.
The action of $D_{3b}$ can be presented as
$$
(H_0, D_{3b} P) =
=
- \sum_j \frac{\partial}{\partial \nu_j} S_{3j}
(H_0, W_j^+ \sum_{m=1}^{\infty} \frac{\tau_j^m}{m!}
\frac{\partial^m}{\partial \mu^m} P)
$$

The last operator $D_4$ gives
$$
(H_0, D_4 P) = - \sum_j \frac{\partial}{\partial \nu_j} (H_0, L_j P)
$$

As the result one can get
$$
\frac{\partial p_0} {\partial t} =
- \sum_j  \frac{\partial}{\partial \nu_j}
(H_0, [L_j + W_j^+ S_{3j} \sum_{m=1}^{\infty} \frac{\tau_j^m}{m!}
\frac{\partial^m}{\partial \mu^m} ] P)
$$
The values in the r.h.s. of the last equation can be interpreted as the
flows $J_j$ along $\nu_j$ axis
$$
J_j =
(H_0, [L_j + W_j^+  S_{3j}  \sum_{m=1}^{\infty} \frac{\tau_j^m}{m!}
\frac{\partial^m}{\partial \mu^m} ] P)
$$
which transfers the last equation in the standard form
\begin{equation}\label{flows}
\frac{\partial p_0} {\partial t} =
- \sum_j  \frac{\partial}{\partial \nu_j}
J_j
\end{equation}

\section{Calculations}

Now we can turn to the direct calculations. The distribution $P$ can be
presented into the following form
$$
P = p_0 H_0 + \sum_{l=0}^{\infty} p_l (p_0) H_l
$$
$$
p_l(p_0) = \sum_{m=1}^{\infty} p_l^{(m)} (p_0)
$$
where the lower index indicates the number of mode and the upper index
in brackets indicates the number of approximation.
As far as we already stated we can fulfill in the external procedure only
one step and it isn't necessary to mark it. The upper index, thus, corresponds
to the internal approximations.

Now we can rewrite expression for $J_j$ in terms of announced decomposition.
Then
$$
J_j =
(H_0, [L_j + W_j^+  S_{3j} \sum_{m=1}^{\infty} \frac{\tau_j^m}{m!}
\frac{\partial^m}{\partial \mu^m} ]
[ p_0 H_0 + \sum_{l=0}^{\infty} p_l (p_0) H_l])
$$
Consider various terms of the last expression. Namely,
$$
(H_0, L_j P) = L_j p_0
$$
Then consider $
(H_0,  W_j^+  S_{3j} \sum_{m=1}^{\infty} \frac{\tau_j^m}{m!}
\frac{\partial^m}{\partial \mu^m}
[ p_0 H_0 + \sum_{l=0}^{\infty} p_l (p_0) H_l])
$
The term $p_0 H_0$ doesn't leads to any
influence because in $ \sum_{m=1}^{\infty} \frac{\tau_j^m}{m!}
\frac{\partial^m}{\partial \mu^m}$ there is at least one operator of the
mode decrease. Then
\begin{eqnarray}
(H_0,  W_j^+  S_{3j}  \sum_{m=1}^{\infty} \frac{\tau_j^m}{m!}
\frac{\partial^m}{\partial \mu^m}
[ p_0 H_0 + \sum_{l=0}^{\infty} p_l (p_0) H_l])
=
(H_0,  W_j^+ S_{3j}  \sum_{m=1}^{\infty} \frac{\tau_j^m}{m!}
\frac{\partial^m}{\partial \mu^m}
 \sum_{l=0}^{\infty} p_l (p_0) H_l    ) =
\nonumber
\\
\nonumber
(H_0, W_j^+ S_{3j} \sum_{i=1}^{\infty} \sum_{m=1}^{\infty} \frac{\tau_j^m}{m!}
\frac{\partial}{\partial \mu^m} p_i(p_0) H_i ) =
W_j^+ S_{3j} \sum_{m=1}^{\infty} \tau_j^m  p_m (p_0) 2^m
\end{eqnarray}

The resulting expression for $J_j$ will be the following
$$
J_j = L_j p_0 + W_j^+ S_{3j} \sum_{m=1}^{\infty} \tau_j^m  p_m (p_0) 2^m
$$
In the main order we can omit $S_{3j}$.  Certainly, the correction term
in $S_{3j}$ has the order greater than $L_j$ has but there is no zero
mode except $L_j p_0$. So, we have to keep $L_j$ and throw away $S_{3j}$
here. Then
$$
J_j = L_j p_0 + W_j^+ \sum_{m=1}^{\infty} \tau_j^m  p_m (p_0) 2^m
$$

Now we can calculate the set of corrections.  The initial condition for
the asymptotic expansion will be the result of the relaxation stage, i.e.
$P_{rel}$. One can easy note that
$$
D_1 P_{rel} = 0
$$
$$
D_2 P_{rel} = 0
$$
$$
D_{3b} P_{rel} = 0
$$
because the fist elementary operator is the mode decrease operator.
In $D_{3a}$ there are only mode increase operators and then
\begin{eqnarray}
D_{3a} P_{rel} = \sum_j  L_j S_{1j} \sum_{l=1}^{\infty} \frac{(-\tau_j)^l}{l!}
(\frac{\partial}{\partial \mu} - 2 \mu )^l p_0 H_0 =
\nonumber
\\
\nonumber
\sum_j  L_j S_{1j} \sum_{l=1}^{\infty} \frac{(-\tau_j)^l}{l!}
(-1)^l p_0 H_l =
\sum_j  L_j S_{1j} \sum_{l=1}^{\infty} \frac{(\tau_j)^l}{l!}
 p_0 H_l
\end{eqnarray}
The last operator contains the small parameter in the first order. In
the main order one can throw $S_{1j}$ out.

Operator $D_4$ has at least the small parameter $r$ in power $2$. So,
it is small in comparison with $D_{3a}$ which has $r$ in the first power.
As far as operators $D_1$, $D_2$ have no action we have to take into account
$D_3$ but we can neglect the action of $D_4$.

As the result we have
$$
p^{(1)}_l = \frac{1}{\lambda_l l!} \sum_j \tau_j^l L_j p_0
$$

In the zero approximation
$$
(\frac{\partial p_0}{\partial t})^{(0)} = 0
$$

The structure of first approximation is rather simple due to the zero
approximation which is localized only at the zero mode. Now the current
approximation has all modes and operator $D_2$ will lead to the non-zero
result. So, the operator $D_2$ will be the main one and one can neglect
$D_3$ and $D_4$.
In $D_2$, thus, one can neglect $S_{2j}$

The time derivative can be also neglected. Really, due  to (\ref{kkk})
$$
(\frac{\partial p_0}{\partial t})^{(1)} \sim
-(H_0 , [D_2+D_3+D_4] H_0 )
$$
in the main order.

One can easy see that
$$
(H_0, D_2 H_0) = 0
$$
$$
(H_0, D_{3b} H_0) = 0
$$
as far as
the first operator in the action of $D_2$ and $D_{3b}$ is the derivative
over $\mu$ which gives zero in application to $H_0 \sim const$.

As far $D_{3a}$ has the small parameter  and $D_4$ has  the result
$(\partial p_0 / \partial t )^{(1)} $  is small.
As far as we calculate in the main order of small parameter  we can neglect
$(\partial p_0 / \partial t )^{(1)}  $.

In all approximations from the second one we can use
(\ref{flows}) and see due to the smallness of
$\partial / \partial \nu_i $ the smallness
of
$(\partial p_0 / \partial t )^{(i)} \ \ i=2 ,3 , 4, ....  $.

We see that the Chapman-Enskog procedure is now reduced to the trivial
equation (except the zero mode)
$$
p^{(i+1)} = D_1^{-1} D_2 p^{(i)}
$$
with the evident initial approximation.  In the main order one can take
$S_{2j}$ away from the operator $D_2$. This form can be obtained from
the simple analysis without Chapman-Enskog formalism.

We can present the last equation in the following form
$$
p_k^{(i+1)} = \sum_{q=1}^{\infty} \Gamma_{kq} p_q^{(i)}
$$
as the matrix representation of the linear operator $D_1^{-1} D_2$ in
the basis $H_i$ (without the zero mode). All $\Gamma_{qq}$ can be put
to zero.

One can note the following important features:
1). Any operator acts until the current moment in
the already fulfilled  part of the external Chapman-Enskog procedure only
one time. 2). Every operator has its own specific structure and can not be
reproduced by the actions of other operators. Both these features allow
to forget about $S_{1j}$, $S_{2j}$, $S_{3j}$. If any operator had been
used more than one time then one would take these operators into account.

The absence of $S_{1j}$, $S_{2j}$, $S_{3j}$ is very important in the context
of the lattice structure of the domain in the $\{ \nu_i \}$ plane. Only
the absence of these operators allows to ignore this lattice structure.

\section{Final expressions}

Now we have to establish the expression for $\Gamma_{kq}$. As far as $D_2$
is the superposition of the mode decrease operators and mode increase
operators it will transfer $H_i$ into superposition of $H_i$. The action
of $D_1^{-1}$ on $H_j$
(except  zero mode) is evident - it is multiplication $H_j$ on
$\lambda_j^{-1}$.

How $D_2$ transfers $H_q$ into $H_k$?
At first according to the definition of $D_2$ the mode decrease operators
will act. The number $l$ of the mode decrease operators will be between
$l_0$ and $q$. Parameter $l_0$ appeared from the evident requirement that
we have to fall lower than the $k$-th level. So, $l_0=1$ if $k>q$ and
$l_0 = q-k+1 $ if $q>k$.  The result of the action of the mode decrease
operators  will give the coefficient $2^l q! / (q-l)!$.     Then
to get $H_k$ one has to apply $k-(q-l)$ mode increase operators which
will give coefficient $(-1)^{k-(q-l)}$. As the result
$$
\Gamma_{kq} =
\sum_j W_j^+
\sum_{l=l_0}^{q}
\frac{\tau_j^{k-q+l}}{(k-q+l)!} \frac{\tau_j^l}{l!} 2^l \frac{q!}{(q-l)!}
$$

Also one can consider another representation of $\Gamma_{kq}$.
At first we fall from the $q$ level to the
$l$ level which gives $\frac{\tau_j^{q-l}}{(q-l)!} \frac{q!}{l!} 2^{q-l}$.
then we increase the mode from the $l$ level to the $k$ level which gives
$\frac{tau_j^{k-l}}{(k-l)!}$ The result will be
$$
\Gamma_{kq} =
\sum_j W_j^+ \sum_{l=0}^{min\{q,k\} =1 }
\frac{\tau_j^{q-l}}{(q-l)!} \frac{q!}{l!} 2^{q-l}
\frac{tau_j^{k-l}}{(k-l)!}
$$

Note that $\Gamma_{kq}$ doesn't depend on the number of approximation.
This will lead to some important consequences.
The first one is the possibility to write equation between $p_k$:
$$
p_k =
\sum_{q=1}^{\infty} \Gamma_{kq} p_q + p^{(1)}_k
$$
or
$$
p_k =
\sum_{q=1}^{\infty} \Gamma_{kq} p_q +
\frac{1}{\lambda_k k!} \sum_j \tau^k_j L_j p_0
$$

Due to the linearity of equations
the decomposition of
the initial approximation will be reproduced in all approximations. Namely,
we shall present $p_l^{(1)}$ in the following form
$$
p_l^{(1)} = \sum_j p_{lj}^{(1)}
$$
where
$$
p_{lj}^{(1)} = \frac{1}{\lambda_l l!} \tau^l_j L_j p_0
$$
and the second  lower index indicates component in the initial approximation
(later $\Gamma_{kq}$ will mixture different components).

The linearity results in the possibility of decomposition
$$
p_k^{(i)} =
\sum_j p_{kj}^{(i)}
$$
Index $j$ has no correspondence here with the direct
decomposition over components.

For $p_{kj}^{(i)}$ the following recurrent expression
$$
p_{kj}^{(i+1)} = \sum_{q=1}^{\infty} \Gamma_{kq} p_{qj}^{(i)}
$$
is valid.

The total amplitude $p_k$ can be also decomposed as
$$
p_k = \sum_j p_{kj}
$$
where $p_{kj}$ are
$$
p_{kj} = \sum_{i=1}^{\infty} p_{kj}^{(i)}
$$

For $p_{kj}$ one can write
$$
p_{kj} = \sum_{q=1}^{\infty} \Gamma_{kq} p_{qj} + p^{(1)}_{kj}
$$

One can also move operators $L_j$ through $\hat{\Gamma} \equiv \{ \Gamma_{kq}
\} $. This gives the seria of relations. For the first approximation one
 can write
$$
p_l^{(1)} = \sum_j L_j a_{lj}^{(1)} p_0
$$
where
$$
a_{lj}^{(1)} = \tau_j^l \frac{1}{\lambda_l l!}
$$
The recurrent relations for $a_{lj}^{(i)}$ will be
$$
a_{kj}^{(i+1)} = \sum_{q=1}^{\infty}  \Gamma_{kq}
a_{qj}^{(i)}
$$

In terms of $a_{kj}^{(i)}$ the value $p_{k}^{(i)}$ can be easily expressed as
$$
p_{k}^{(i)} = \sum_j a_{kj}^{(i)} L_j p_0
$$

Having introduced
$$
a_{kj} = \sum_{i=1}^{\infty} a_{qj}^{(i)}
$$
one can get
$$
a_{kj} = \sum_{q=1}^{\infty}  \Gamma_{kq}
a_{qj}
+ a_{kj}^{(1)}
$$

In terms of $a_{kj}$ the value $p_{k}$ can be easily expressed as
$$
p_{k} = \sum_j a_{kj} L_j p_0
$$

The given decompositions are rather attractive from the first point of
view, but actually one can not simplify the problem with the help of these
decompositions.

The equation for $p_0$ will the following one
$$
\frac{\partial p_0}{\partial t} =
-\sum_j \frac{\partial}{\partial \nu_j}
\sum_i
(\delta_{ij} +
W_j^+ S_{3j} \sum_{m=1}^{\infty}  \tau^m_j 2^m  a_{mi} ) L_i p_0
$$
and in the main order
$$
\frac{\partial p_0}{\partial t} =
-\sum_j \frac{\partial}{\partial \nu_j}
\sum_i
(\delta_{ij} +
W_j^+  \sum_{m=1}^{\infty}  \tau^m_j 2^m  a_{mi} ) L_i p_0
$$
where indexes $i$ and $j$ marks components. The last equation has the
standard form investigated in \cite{Kuni-mul}. The part of \cite{Djik}
concerning the solution of the last equation in the two dimensional
sense
is also acceptable as far as
it reproduces \cite{TechPhys} even in details\footnote{except
several misprints appeared in \cite{Djik}. Also the initial matrix (with
dimension  $2$)  isn't   diagonal   which   doesn't   produce   any
difficulties.}.

Now we shall present another
method to calculate correction terms in the Chapman-Enskog procedure.

One can easily note that recurrent equations  for $p_l^{(i)}$ and
 for $p_{l \ j }^{(i)}$
will lead to
$$
\lim_{i \rightarrow \infty} p_{l \ j }^{(i)} \rightarrow const_{l\ j}
 (\gamma_{max})^i
$$
$$
\lim_{i \rightarrow \infty} p_{l}^{(i)} \rightarrow const_l (\gamma_{max})^i
$$
where
$\gamma_{max}$ is the eigenvalue of $\hat{\Gamma} \equiv \{ \Gamma_{pq} \}$
with the maximal absolute value.

Then one can say that starting from some number $m_{lim}$ the tails of
sums $\sum_i p_{l\ j}^{(i)}$, $\sum_i p_{l}^{(i)}$ resemble the tail
of  geometric progressions. Then one can easily calculate these sums
$$
\sum_{i=m_{lim}}^{\infty}
 p_{l\ j}^{(i)} \sim \frac{p^{(m_{lim})}_{l\ j}}{1-\gamma_{max}}
$$
and
$$
\sum_{i=m_{lim}}^{\infty}
 p_{l}^{(i)} \sim \frac{p^{(m_{lim})}_l}{1-\gamma_{max}}
$$
The first $m_{lim}$ terms have to be calculated explicitly. The boundary
$m_{lim}$ which depends on $j,l$ can be found as the characteristic boundary
when $\hat{\Gamma}^{i+1} / \hat{\Gamma}^i$ approaches some constant
independent
on $i$. This procedure gives also the value of $\gamma_{max}$.

Certainly, the most interesting situation is the strong manifestation of the
thermal effects. Here the  thermal effects can not be considered
as some corrections but radically change the character of the process.

 Really, the main dependence of the nucleation rate is accumulated
in exponent of the critical embryo free energy. The giant renormalization
due to the thermal effects means that the temperature of the critical
embryo really differs from the temperature of the vapor-gas media. The
relative
difference of the temperature expressed in the units of $\beta$ (in estimates
we can forget about different components) can attain several units. It
means that the characteristic value
$\mu_0$ of $\mu$ attains several units.

One can easily note that to reproduce the real solution (i.e. to attain
the values of $\mu$ in several units) we have to take into account
a great number of
modes.
Really, the Hermite polynomials $H_n$ are the polynomials of power $n$.
The characteristic region of localization of the function $H_n (\mu)
\exp(-\mu^2)$ is $[ \sim - \sqrt{n} , \sim \sqrt{n} ]$. So, to describe the
situation correctly we have to take into account at least $\mu_0^2$ modes.
This quantity  equals to the dimension of the matrix $\hat{\Gamma}$ which
becomes also great. This produces numerical difficulties.

Consider the matrix $\hat{\Gamma}$.
Due to factorials in the denominators
the limit of  the elements with a big indexes is going to zero. It takes
place
under the arbitrary $\tau_j$. The r.h.s. of the matrix equation,
i.e. the known vector
$p^{(1)} = \{ p^{(0)}_i \}$ also has vanishing elements when $i \rightarrow
\infty$.
The structure of the matrix $\hat{\Gamma}$ is the following: the upper
triangle matrix has the elements with big values due to the big coefficient
appeared from the action of the mode decrease. The lower triangle matrix
is "smaller" than the upper one.
Thus, the following method will be rather effective\footnote{With the special
account of the Chapman-Enskog specific features}: one can split matrix
$\hat{\Gamma}$ into the upper triangle matrix (with big elements) and
the lower triangle matrix (with moderate elements). Then at every step
of the iteration procedure the upper triangle matrix will be inverted
(it is easy to do). The zero mode (in fact it is the first line in the
matrix equation) will be calculated explicitly.

One can easily note
the the iteration solutions of the matrix equation return us to the
initial formulation of the problem at the level of the recurrent relations.
So, the presented approach to calculate the maximal eigenvalue
 gets now a solid ground. It is also
more effective because it allows to estimate the necessary number of modes
taken into account directly (by the smallness of the higher mode at every
step).

\pagebreak

\end{document}